\documentclass[twocolumn]{aastex631}

\received{Oct. 31, 2024}
\accepted{May 15, 2024}

\submitjournal{one of the AAS Journals}
\shorttitle{Oscillations in RVs of \target\ with NEID}
\shortauthors{Luhn et al.}

\begin{document}
\newcommand{\aastex}{AAS\TeX}

\newcommand{\PSUAA}{Department of Astronomy \& Astrophysics, 525 Davey Laboratory, 251 Pollock Road, Penn State, University Park, PA, 16802, USA}
\newcommand{\PSUCEHW}{Center for Exoplanets and Habitable Worlds, 525 Davey Laboratory, 251 Pollock Road, Penn State, University Park, PA, 16802, USA}
\newcommand{\PSETI}{Penn State Extraterrestrial Intelligence Center, 525 Davey Laboratory, 251 Pollock Road, Penn State, University Park, PA, 16802, USA}
\newcommand{\UA}{Steward Observatory, The University of Arizona, 933 N.\ Cherry Ave, Tucson, AZ 85721, USA}
\newcommand{\UAA}{Department of Astronomy and Steward Observatory, University of Arizona, Tucson, AZ 85721, USA}
\newcommand{\Penn}{Department of Physics and Astronomy, University of Pennsylvania, 209 S 33rd St, Philadelphia, PA 19104, USA}
\newcommand{\Caltech}{Department of Astronomy, California Institute of Technology, Pasadena, CA 91125, USA}
\newcommand{\STScI}{Space Telescope Science Institute, 3700 San Martin Dr, Baltimore, MD 21218, USA}
\newcommand{\JHU}{Department of Physics and Astronomy, Johns Hopkins University, 3400 N Charles St, Baltimore, MD 21218, USA}
\newcommand{\GoddardESAL}{Exoplanets and Stellar Astrophysics Laboratory, NASA Goddard Space Flight Center, Greenbelt, MD 20771, USA}
\newcommand{\GoddardISTD}{Instrument Systems and Technology Division, NASA Goddard Space Flight Center, Greenbelt, MD 20771, USA}
\newcommand{\GSFC}{NASA Goddard Space Flight Center, Greenbelt, MD 20771, USA}
\newcommand{\NOAO}{U.S. National Science Foundation National Optical-Infrared Astronomy Research Laboratory, 950 N.\ Cherry Ave., Tucson, AZ 85719, USA}
\newcommand{\UW}{Wisconsin address goes here}
\newcommand{\MacquarieSchool}{School of Mathematical and Physical Sciences, Macquarie University, Balaclava Road, North Ryde, NSW 2109, Australia}
\newcommand{\MacquarieCentre}{Astrophysics and Space Technologies Research Centre, Macquarie University, Balaclava Road, North Ryde, NSW 2109, Australia}
\newcommand{\MacquarieAAO}{Australian Astronomical Optics, Macquarie University, Balaclava Road, North Ryde, NSW 2109, Australia}
\newcommand{\NIST}{National Institute of Standards \& Technology, 325 Broadway, Boulder, CO 80305, USA}
\newcommand{\CUBoulder}{Department of Physics, 390 UCB, University of Colorado, Boulder, CO 80309, USA}
\newcommand{\JPL}{Jet Propulsion Laboratory, California Institute of Technology, 4800 Oak Grove Drive, Pasadena, California 91109}
\newcommand{\MIT}{Kavli Institute for Astrophysics and Space Research, Massachusetts Institute of Technology, Cambridge, MA, USA}
\newcommand{\UCI}{Department of Physics \& Astronomy, The University of California, Irvine, Irvine, CA 92697, USA}
\newcommand{\Carleton}{Carleton College, One North College St., Northfield, MN 55057, USA}
\newcommand{\Carnegie}{Earth and Planets Laboratory, Carnegie Institution for Science, 5241 Broad Branch Road, NW, Washington, DC 20015, USA}
\newcommand{\PSUICS}{Institute for Computational and Data Sciences, The Pennsylvania State University, University Park, PA, 16802, USA}
\newcommand{\PSUCASt}{Center for Astrostatistics, 525 Davey Laboratory, 251 Pollock Road, Penn State, University Park, PA, 16802, USA}
\newcommand{\NESSF}{NASA Earth and Space Science Fellow}
\newcommand{\Princeton}{Department of Astrophysical Sciences, Princeton University, 4 Ivy Lane, Princeton, NJ 08540, USA}
\newcommand{\RUSSELL}{Henry Norris Russell Fellow}
\newcommand{\IAS}{Institute for Advance Study, 1 Einstein Drive, Princeton, NJ 08540, USA}
\newcommand{\Tsinghua}{Department of Astronomy, Tsinghua University, Beijing 100084, China}
\newcommand{\FlatironCCA}{Center for Computational Astrophysics, Flatiron Institute, 162 Fifth Avenue, New York, NY 10010, USA}
\newcommand{\ETH}{ETH Zurich, Institute for Particle Physics \& Astrophysics, Zurich, Switzerland}
\newcommand{\UMinn}{Department of Physics \& Astronomy, University of Minnesota Duluth, Duluth, MN 55812, USA}
\newcommand{\UIUC}{Department of Astronomy, University of Illinois at Urbana-Champaign, Urbana, IL 61801, USA}
\newcommand{\TIFR}{Department of Astronomy and Astrophysics, Tata Institute of Fundamental Research, Homi Bhabha Road, Colaba, Mumbai 400005, India}

\newcommand{\API}{Anton Pannekoek Institute for Astronomy, University of Amsterdam, Science Park 904, 1098 XH Amsterdam, The Netherlands}
\newcommand{\target}{HD\,142091}

\newcommand{\USyd}{Sydney Institute for Astronomy, School of Physics, University of Sydney, Sydney, NSW 2006, Australia}

\newcommand{\Teff}{T_{\mathrm{eff}}}
\newcommand{\logg}{\log{g}}
\newcommand{\vsini}{v \sin{i}}
\newcommand{\logRHK}{\log{R'_{\mathrm{HK}}}}
\newcommand{\SHK}{$S_{\mathrm{HK}}$}
\newcommand{\ms}{m$\,$s$^{-1}$}
\newcommand{\cms}{cm$\,$s$^{-1}$}
\newcommand{\kms}{km$\,$s$^{-1}$}
\newcommand{\Ha}{H$\alpha$}
\newcommand{\Kepler}{\emph{Kepler}}
\renewcommand*{\sectionautorefname}{Section}
\renewcommand*{\subsectionautorefname}{Section}
\renewcommand*{\subsubsectionautorefname}{Section}

\def\app#1#2{%
  \mathrel{%
    \setbox0=\hbox{$#1\sim$}%
    \setbox2=\hbox{%
      \rlap{\hbox{$#1\propto$}}%
      \lower1.1\ht0\box0%
    }%
    \raise0.25\ht2\box2%
  }%
}
\def\approxprop{\mathpalette\app\relax}

\newcommand{\sgnpz}{\mathrm{s}}

\definecolor{m1}{RGB}{0, 162, 255}
\definecolor{m2}{RGB}{0, 118, 186}
\definecolor{m3}{RGB}{0, 171, 142}
\definecolor{m4}{RGB}{29, 177, 0}
\definecolor{m5}{RGB}{1, 113, 0}

\definecolor{jkl}{rgb}{0.4, 0.0, 0.6}
\newcommand{\jkl}[1]{\textcolor{jkl}{\bf [JKL: #1]}}
\newcommand{\todo}[1]{\textcolor{jkl}{\bf [to do: #1]}}

\title{Time-resolved p-mode oscillations for subgiant HD 142091 with NEID at WIYN}

\correspondingauthor{Jacob K. Luhn}
\email{jacob.luhn@jpl.nasa.gov}

\author[0000-0002-4927-9925]{Jacob K. Luhn}
\altaffiliation{NASA Postdoctoral Fellow}
\affil{\UCI}
\affil{\JPL}

\author[0000-0003-0149-9678]{Paul Robertson}
\affil{\UCI}

\author[0000-0003-1312-9391]{Samuel Halverson}
\affil{\JPL}

\author[0000-0002-5463-9980]{Arvind F.\ Gupta}
\affil{\NOAO}

\author[0000-0002-9337-0902]{Jared C. Siegel}
\affil{\Princeton}

\author[0000-0001-6160-5888]{Jason T.\ Wright}
\affil{\PSUAA}
\affil{\PSUCEHW}
\affil{\PSETI}

\author[0000-0001-6545-639X]{Eric B.\ Ford}
\affil{\PSUAA}
\affil{\PSUCEHW}
\affil{\PSUICS}
\affil{\PSUCASt}

\author[0000-0001-9596-7983]{Suvrath Mahadevan}
\affil{\PSUAA}
\affil{\PSUCEHW}

\author[0000-0001-5222-4661]{Timothy R. Bedding}
\affil{\USyd}


\author[0000-0003-0353-9741]{Jaime A. Alvarado-Montes}
\affil{\MacquarieAAO}
\affil{\MacquarieCentre}

\author[0000-0003-4384-7220]{Chad F.\ Bender}
\affil{\UA}

\author[0000-0002-3610-6953]{Jiayin Dong}
\affil{\FlatironCCA}
\affil{\UIUC}

\author[0000-0002-1664-3102]{Fred Hearty}
\affil{\PSUAA}
\affil{\PSUCEHW}

\author[0000-0002-9632-9382]{Sarah E.\ Logsdon}
\affil{\NOAO}

\author[0000-0002-0048-2586]{Andrew Monson}
\affil{\UA}

\author[0000-0003-0241-8956]{Michael W.\ McElwain}
\affil{\GoddardESAL} 

\author[0000-0001-8720-5612]{Joe P.\ Ninan}
\affil{\TIFR}

\author[0000-0002-2488-7123]{Jayadev Rajagopal}
\affil{\NOAO}

\author[0000-0001-8127-5775]{Arpita Roy}
\affiliation{Astrophysics \& Space Institute, Schmidt Sciences, New York, NY 10011, USA}

\author[0000-0002-4046-987X]{Christian Schwab}
\affil{\MacquarieSchool}
\affil{\MacquarieCentre}

\author[0000-0001-7409-5688]{Gudmundur Stefansson}
\affil{\API}

\author[0000-0002-5951-8328]{Daniel J. Stevens}
\affil{\UMinn}

\author[0000-0002-4788-8858]{Ryan C. Terrien}
\affil{\Carleton}

\author[0000-0002-6937-9034]{Sharon Xuesong Wang}
\affiliation{\Tsinghua}

\author[0000-0001-5290-2952]{Jinglin Zhao}
\affiliation{DTU Space, Elektrovej, Building 328, DK-2800 Kgs. Lyngby, Denmark}
\affil{\PSUAA}

\begin{abstract}
Detections of Earth-analog planets in radial velocity observations are limited by stellar astrophysical variability occurring on a variety of timescales. Current state-of-the-art methods to disentangle potential planet signals from intrinsic stellar signals assume that stellar signals introduce asymmetries to the line profiles that can therefore be separated from the pure translational Doppler shifts of planets. Here, we examine this assumption using a time series of resolved stellar p-mode oscillations in \target\ ($\kappa$~CrB), as observed on a single night with the NEID spectrograph at 2-minute cadence and with 25~\cms\ precision. As an evolved subgiant star, this target has p-mode oscillations that are larger in amplitude ($4$--$8$~\ms) and occur on longer timescales (80~min.) than those of typical Sun-like stars of RV surveys, magnifying their corresponding effects on the stellar spectral profile. We show that for \target, p-mode oscillations manifest primarily as pure Doppler shifts in theaverage line profile---measured by the cross correlation function (CCF)---with ``shape-driven" CCF variations as a higher-order effect. Specifically, we find that the amplitude of the shift varies across the CCF bisector, with 10\% larger oscillation amplitudes closer to the core of the CCF, and 25\% smaller oscillation amplitudes for bisector velocities derived near the wings; we attribute this trend to larger oscillation velocities higher in the stellar atmosphere. Using a line-by-line analysis, we verify that a similar trend is seen as a function of average line depth, with deeper lines showing larger oscillation amplitudes. Finally, we find no evidence that p-mode oscillations have a chromatic dependence across the NEID bandpass beyond that due to intrinsic line depth differences across the spectrum.

\end{abstract}
  

\section{Introduction}
Current optical radial velocity (RV) spectrographs ESPRESSO \citep{Pepe2021}, EXPRES \citep{Petersburg2020}, NEID \citep{Halverson2016}, KPF \citep{Gibson2016}, MAROON-X \citep{Seifahrt2018} have successfully pushed the instrumental RV precision floor to the 30~\cms\ level. However, despite the improvements offered by these instruments, low-amplitude exoplanet signals (i.e., Earth analogs orbiting in the habitable zones of Sun-like stars) remain difficult to disentangle from intrinsic stellar sources of RV variability, which operate at amplitudes as small as 0.3 to well over 1~\ms\ \citep[e.g.,][]{Fischer2016,Luhn2020a}.

The astrophysical sources of RV variability distort the average stellar line profiles, different from the bulk center-of-mass shifts expected from planets \citep[e.g.,][]{Dravins1981,Saar1997,Telting1997,Meunier2010b}. With the stability of current extreme precision RV (EPRV) spectrographs, we can hope to distinguish such ``shape-driven'' line changes from the ``shift-driven'' line changes of exoplanets. Several previous studies have explored various computational methods, including using the autocorrelation of the cross-correlation function (CCF)---a transformation that is invariant to pure shift changes \citep{CollierCameron2021}; parameterizing the CCF shapes and characterising the shift-invariant changes of the CCF in the Fourier domain \citep{zhao_fiesta_2022}; as well as several machine-learning approaches trained on both simulated and solar data sets \citep{deBeurs2022,Colwell2023}. However, a recent comparison of several stellar variability mitigation strategies showed that existing methods do not agree for any given star, and are inconsistent from star to star \citep{Zhao2022}.

Here, we focus on a specific source of astrophysical variability: stellar p-mode oscillations, the convection-driven pressure waves that travel through the interior of a star. These oscillations occur with characteristic timescale and amplitude set by the temperature, mass, and radius of the star and are the foundation of asteroseismology \citep{Chaplin2013}. In photometry, oscillations arise from the small brightness fluctuations as expanding (cooling) and contracting (heating) material changes temperature. In RV observations, these effects are even more prominent, as the measurements directly probe the velocities of the stellar surface. For the Sun, p-mode oscillations occur with maximum power at a cycle period of 5.5 minutes and peak-to-peak RV amplitude of 2~\ms\ \citep{Kjeldsen2008, fredslund_andersen_oscillations_2019}. For Sun-like stars, the common approach to mitigate the effect of p-modes is to expose for one p-mode cycle (or sometimes for an integer number of p-mode cycles; \citealt{Chaplin2019}), which can reduce the residual RMS to the 10--20~\cms\ level. Below the 10-\cms\ level, or for large telescopes where such long exposure times ($\sim$11 minutes) require many back-to-back exposures, this technique becomes less efficient. In addition, the observed oscillations are the superposition of many modes within a range of periods, and it is an approximation to characterize them by a single period.

Building on the approach of the ``shape-vs-shift'' techniques described above, we seek to characterize the spectral effects of p-mode oscillations on stellar line profiles. To resolve p-mode effects in the time domain, we turn to a more evolved subgiant star, whose larger convective power yields longer cycle periods ($> 1$ hour) and amplitudes (several \ms). With such well resolved oscillations on a stellar target, we can investigate how p-mode oscillations affect line shape morphology, and assess the applicability of leading ``shape-vs-shift'' approaches to reduce their effects in RV time series. Previous models of p-mode oscillations predict that radial p-modes produce pure shifts, whereas nonradial p-modes in rotating stars create shape changes \citep{Telting1997}, which was confirmed observationally in evolved stars \citep{Hekker2010}. These previous studies focused on characterizing the line profiles of individual modes. In this study, we wish to assess the line profile characteristics as would be seen over a few oscillation cycles, without resolving the individual mode frequencies. Such an approach can more directly reveal how ``shape-vs-shift'' mitigation techniques may hope to improve over the \citet{Chaplin2019} exposure-time-binning technique. If CCF asymmetries are detectable in RVs without the necessary time baseline to resolve individual modes, these techniques could provide additional improvement by disentangling any remaining CCF asymmetries after exposure time binning, or via an observational approach that samples the oscillations rather than averaging over them.

The manuscript is structured as follows: In \autoref{sec:target}, we describe the target for our observations, \target, and our observation strategy. In \autoref{sec:observations}, we describe the NEID instrument and our observations. In \autoref{sec:analysis}, we investigate a number of different line-variability metrics to characterize the p-mode oscillation time series. \autoref{sec:discussion} discusses the implications of the results as applied to EPRV surveys and mitigation techniques. Finally, in \autoref{sec:conclusions}, we summarize our findings.
    
\section{\target}\label{sec:target}
\target\ ($\kappa$~CrB) is a subgiant star and known host to a long-period ($P=1254$~d) exoplanet \citep{Johnson2008}; its stellar properties are given in \autoref{tbl:stellar_properties}, which come from the NASA Exoplanet Archive\footnote{\url{https://exoplanetarchive.ipac.caltech.edu}}. Specifically, the derived stellar parameters are adopted from \citet{Teng2023}, who used the \texttt{isoclassify} \citep{Huber2017,Berger2020,Berger2023} to perform a Bayesian estimate with stellar isochrones. \texttt{isoclassify} first derives a luminosity using the Gaia parallax and estimated stellar properties from spectroscopy; the final stellar parameters come from a direct integration of MESA isochrones. As a reference point, we use the stellar properties (surface gravity and effective temperature) along with the relations given in \citet{Luhn2023} \citep[which themselves build upon the scaling relations in ][]{Kjeldsen1995,Kallinger2014,Chaplin2019,Guo2022}. We calculate the estimated p-mode oscillation RMS amplitude (3.27~\ms) and p-mode period (84~min.) for this star; these numbers are consistent with previous Keck-HIRES observations presented in \citet{Luhn2020a} for this target, which show oscillations resolved in the time domain with 1.5~hr period and amplitude of roughly 3-\ms\ on two different nights.

Similar to these Keck-HIRES observations, our strategy is to resolve p-mode oscillations on \target\ over the course of a single night, aided by the higher precision and instrument stability offered by NEID.

\begin{deluxetable*}{lccc}
\tablecaption{Summary of Stellar Parameters for HD 142091 \label{tbl:stellar_properties}}
\tablehead{\colhead{~~~Parameter}&  \colhead{Value}&
\colhead{Description}&
\colhead{Reference}}
\startdata
\multicolumn{4}{l}{\hspace{-0.2cm}Alternate Identifiers:}  \\
~~~TIC & 334753043 & TESS Input Catalog & Stassun \\
~~~Gaia DR3 &  1372702716380418688 & Gaia & Gaia DR3\\
~~~HIP & 77655 & HIPPARCOS Catalog  &  HIPPARCOS \\
\multicolumn{4}{l}{\hspace{-0.2cm} Coordinates and Parallax:} \\
~~~$\alpha_{\mathrm{J2000}}$ &    15:51:13.93 & Right Ascension (RA) & Gaia DR3\\
~~~$\delta_{\mathrm{J2000}}$ &   +35:39:26.57 & Declination (Dec) & Gaia DR3\\
~~~$\varpi$  & $33.3433 \pm 0.08$ &  Parallax (mas) & Gaia DR3 \\
\multicolumn{4}{l}{\hspace{-0.2cm} Broadband photometry:}  \\
~~~TESS &  $3.9341 \pm 0.0075$  & TESS & Stassun \\
~~~$G$  &  $4.5113 \pm 0.0029]$ & Gaia & Gaia DR3\\
~~~$V$ & $4.79 \pm 0.023$ & TESS & Stassun\\
~~~$B_p$  &  $5.0258 \pm 0.0006$ & Gaia & Gaia DR3 \\
~~~$R_p$  &  $3.8582 \pm 0.0023$ & Gaia & Gaia DR3 \\
\multicolumn{4}{l}{\hspace{-0.2cm} Derived Stellar Parameters:}\\
~~~$v\sin{i}$ & $1.21$ & Rotational velocity (\kms) &  Sato\\
~~~$T_{\mathrm{eff}}$ & $4840$ & Effective Temperature (K) &  Teng\\
~~~$L_\star$ & $10.86^{+4.38}_{-3.18}$ &  Luminosity (${\rm L}_\odot$) & Teng\\
~~~$R_\star$ & $4.68^{+0.84}_{-0.71}$ & Stellar Radius (${\rm R}_\odot$)   &  Teng\\
~~~$M_\star$ & $1.33\pm0.25$ & Stellar Mass (${\rm M}_\odot$)  &  Teng\\
~~~$\log g$ & $3.20\pm0.1$ & Surface Gravity ($\log$ (g cm$^{-3}$)) &  Teng\\
\enddata
\tablenotetext{}{References are: HIPPARCOS \citep{ESA1997}, Stassun \citep{Stassun2019}, Gaia DR3 \citep{GaiaCollaboration2022}, Sato \citep{Sato2012}, Teng \citep{Teng2023}}

\end{deluxetable*}

\section{NEID Observations}\label{sec:observations}
\begin{figure}
    \centering
    \includegraphics[width=\columnwidth]{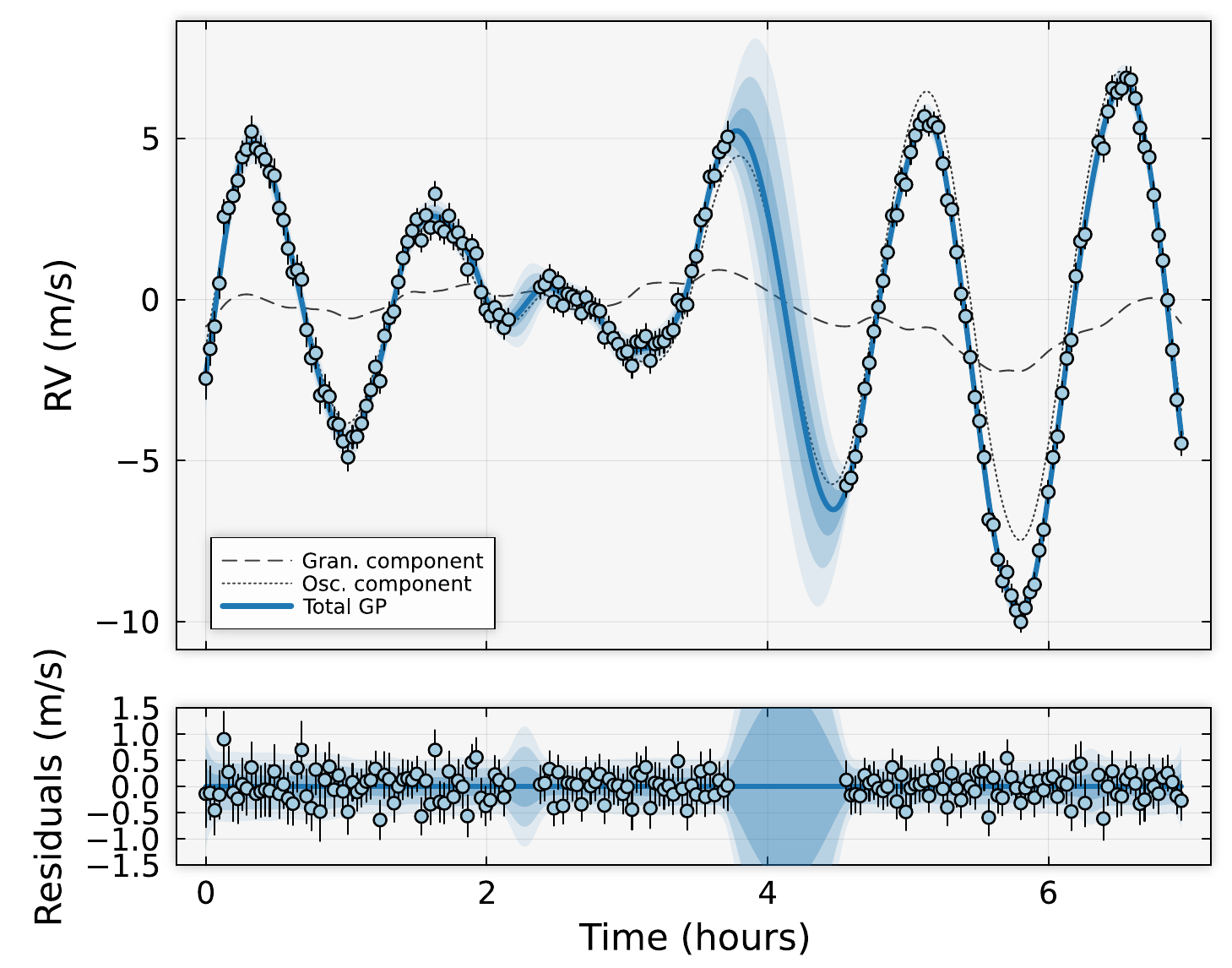}
    \caption{NEID RV time series of \target\ on April 25, 2022, in which we have resolved p-mode oscillations. Gaps in the time series result from a combination of scheduled intermediate calibrations, acquisition errors, and a VNC crash during the night, as described in the text. The blue line and ribbons shows a GP fit to the time series and the 1-, 2-, and 3-$\sigma$ uncertainties using a two-component GP model that includes granulation and oscillation \citep{Luhn2023,Gupta2022}. The decomposed oscillation and granulation components of the GP model are shown in dotted and dashed lines, respectively. The residuals about the fit are shown in the bottom panel and have an RMS of 25~\cms.}
    \label{fig:neid_ts}
\end{figure}

NEID is an ultra-stabilized, high resolution ($R\sim$115,000) Doppler spectrograph located at the 3.5-m WIYN telescope\footnote{The WIYN Observatory is a joint facility of the University of Wisconsin-Madison, Indiana University, the National Optical Astronomy Observatory and the University of Missouri.} at Kitt Peak National Observatory in Arizona \citep{Schwab2016}. It operates in a wide wavelength range covering 380--930~nm, and has demonstrated $<1$~\ms\ long-term precision on stellar targets \citep[e.g.,][]{Gupta2024}. Our NEID observations of \target\ were collected over the course of a single night, 2022 April 25, with observations occurring with 2-minute cadence (90~s exposures + 27~s read out). NEID obtained 191 spectra during the night, with a typical signal-to-noise ratio (SNR) of 226 at 5500~\AA, which were processed with the standard NEID Data Reduction Pipeline (DRP) version 1.1.3 to extract RVs with typical uncertainty 37~\cms; the RV time series is shown in \autoref{fig:neid_ts}. During the course of the night, observations were interrupted for the NEID nightly intermediate calibration sequence (seen near the 2-hour mark in \autoref{fig:neid_ts}). Later in the night, a virtual network computing (VNC) crash led to loss of/suboptimal guiding and a drop in SNR for several observations; this gap in observations was lengthened by acquisition errors near zenith once the VNC connection was reestablished. Together, the VNC crash and zenith acquisition errors account for the large gap in the time series seen near the 4-hour mark. One final observation suffered from a loss of guiding toward the end of the time series (near the 6-hour mark), but was terminated early in its exposure. In total, we discarded 2 terminated observations (exposed for 8 and 11~s instead of nominal 90~s exposures), and 8 loss-of-guiding observations (as indicated by negative SNR values)---the final time series is 181 observations.

Despite these gaps in the time series, p-mode oscillations were cleanly resolved with NEID. We follow the procedure demonstrated for the observations of HD 35833 in \S~4.3.1 of \citet{Gupta2022} and condition the time series with a two-component Gaussian process (GP) model with fixed hyperparameters from the kernels in \citet{Luhn2023} for granulation and oscillations. The residuals about the GP posterior have RMS $<$ 25~\cms, 31\% lower than the median 37~\cms\ single measurement precision estimated by the NEID DRP. To test whether the GP fitting has absorbed any white noise, we generated 500 time series drawn from our GP models and added 37~\cms\ white noise; our recovered GP fits to the simulated time series have a mean RMS residual of $32.3\pm1.9$~\cms, indicating that the GP fitting procedure can be expected to absorb $\sim$10\% of the white noise. Returning to the observed time series, the residual RMS to the GP fit of our data suggest that the individual errors from the DRP are therefore likely overestimated, and that a single measurement precision near 28~\cms\ is more consistent with our observations.

\section{RV effects of p-mode oscillations}\label{sec:analysis}

Because the RV time series is dominated by p-mode oscillations, we can perform a variety of analyses to investigate their effect on RV spectra and associated data products.  

\subsection{CCF line profiles}
We begin with an investigation of the line morphology, as measured by the cross correlation function (CCF) between each target spectrum and a numerical mask \citep{Baranne1996,Pepe2002}, which for \target\ is the K2 ESPRESSO mask. The default NEID pipeline calculates and produces order-by-order CCFs (echelle orders 58–163, $\sim$380 nm -- 1.1$\mu$m) with a velocity array spanning $-100$ to $100$~\kms\ in increments of 0.25~\kms, along with each order's weighting (given by header keywords `\texttt{CCFNNN}' in extension 12 of the L2 files) in the summation of the cumulative CCF used to derive the bulk RV (`\texttt{CCFRVMOD}'). After summing the weighted CCF of each order, we noticed a systematic linear slope across the full CCF velocity span (which can arise, e.g., from edge effects when many lines are located near the edge of a spectral order, depending on a star's systemic velocity), which we subtract by fitting a line to the CCF points with velocities $|v_{CCF}| > 30$~\kms. Finally, we normalize the CCF of each observation by dividing by the 99th percentile value.

\subsubsection{CCF morphology and residuals}

\begin{figure}
    \centering
    \includegraphics[width=\columnwidth]{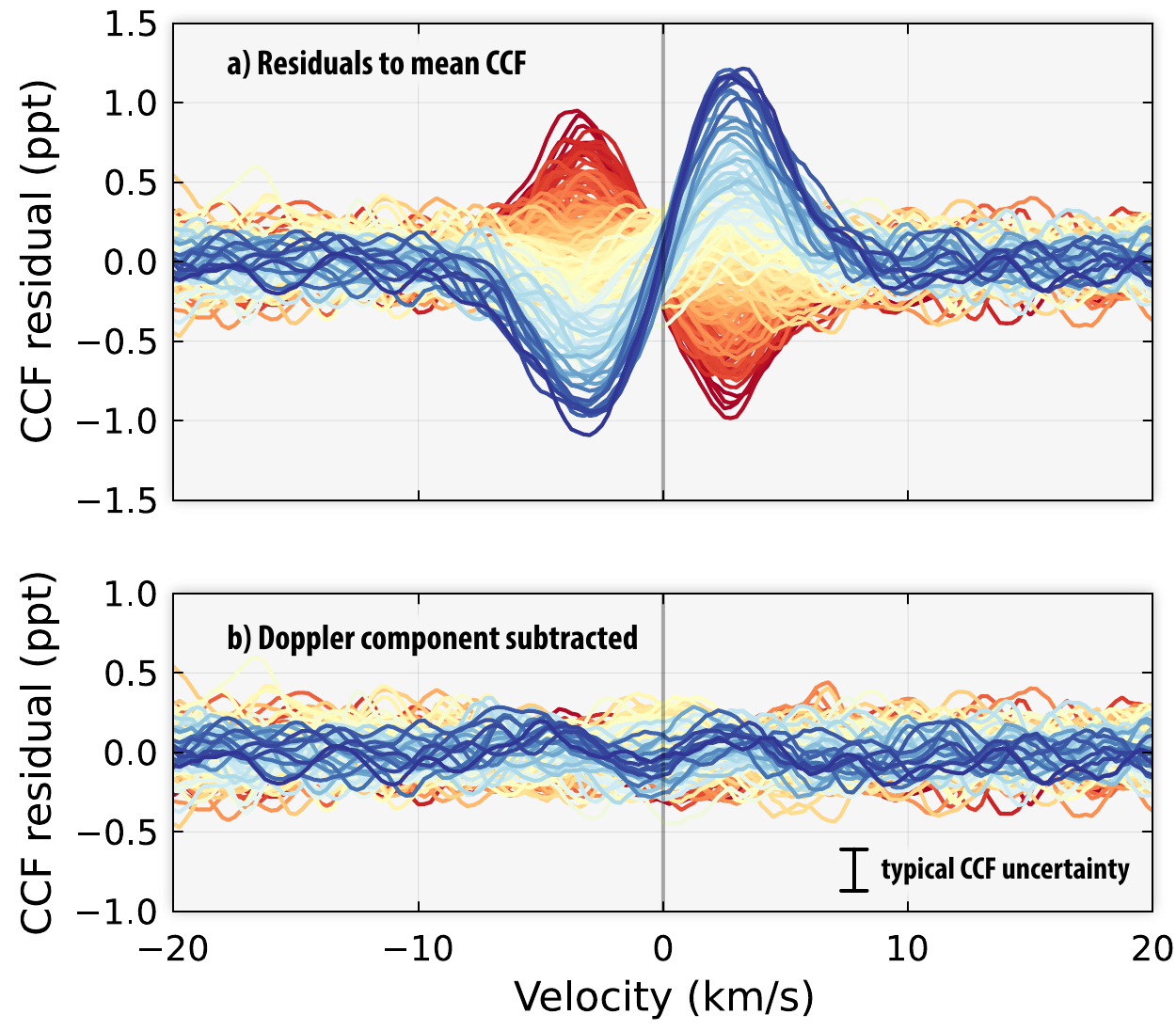}
    \caption{\emph{Top:} Boxcar-smoothed CCF residuals for NEID, with each observation color coded by the spectrum's bulk RV. The CCF residuals are largely antisymmetric about 0, indicating CCF changes that behave like pure Doppler shifts. \emph{Bottom:} CCF residuals from the top panel after removing the pure Doppler shift component expected based on the pipeline RV. This panel is equivalent to subtracting the template CCF Doppler shifted by each observation's RV from the raw CCFs directly. Some additional structure is hinted at, indicating possible higher-order effects not due to pure shifts; these are near or within the typical noise floor of the CCFs.}
    \label{fig:ccf_residuals}
\end{figure}

To identify changes in CCF shape from observation to observation, we construct a mean CCF (our ``template" CCF) by computing the unweighted average across all observations, and compute the residuals of each observation's CCF to the template CCF. These residuals can be seen in \autoref{fig:ccf_residuals}, where we have also performed a 5-point (1.25~\kms) boxcar smoothing of the CCFs and the template prior to calculating the residuals. The CCF residuals in \autoref{fig:ccf_residuals} are largely antisymmetric about 0, behavior that closely follows that of pure translational Doppler shifts.

\begin{figure*}
    \centering
    \includegraphics[width=\textwidth]{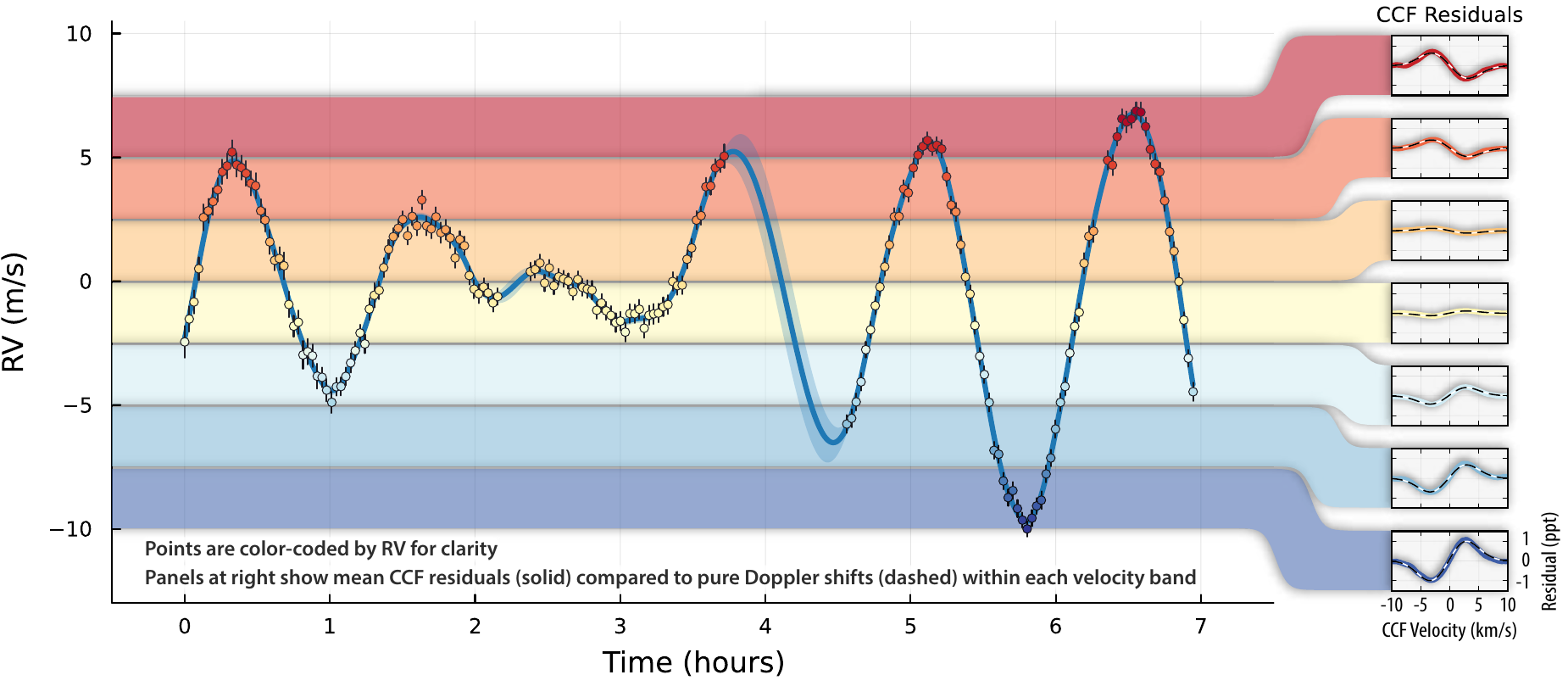}
    \caption{NEID time series with CCF residuals in 7 RV bins. The time series is color coded by RV for additional clarity. The insets show the mean CCF residual (solid, colored lines) compared to pure Doppler shifts (black and white dashed lines) within each velocity band. The CCF residuals closely match those of pure Doppler shifts, indicating primarily shift-driven, rather than shape-driven RV variations.}
    \label{fig:ccf_residuals_ts}
\end{figure*}

To draw a closer comparison to pure Doppler shifts, we bin the time series into 7 RV bins, each with width of 2.5~\ms. We then compute the average CCF for all observations within a given RV bin. We compare the mean residuals (using the same template CCF as above) in each bin to pure Doppler shifts (calculated by translationally shifting the template CCF). \autoref{fig:ccf_residuals_ts} shows the time series and RV bins used along with the resulting residuals for each bin as compared to pure Doppler shifts. We find that the CCF residuals can be explained almost fully by pure Doppler shifts (the median ratio of the RMS within $\pm$10~\kms\ to the RMS outside $\pm$10~\kms\ is 1.069), indicating that the p-mode oscillations in \target\ produce RVs that are primarily shift-driven rather than shape-driven. In the context of previous studies \citep{Telting1997,Hekker2010}, this result is consistent with the fact that the p-mode oscillations we observe in this short observing sequence are dominated by modes with low angular degree ($\ell=0$, 1 and 2).

\subsubsection{CCF Bisector Velocities}\label{sec:bis_velocities}

\begin{figure*}
    \centering
    \includegraphics[width=\textwidth]{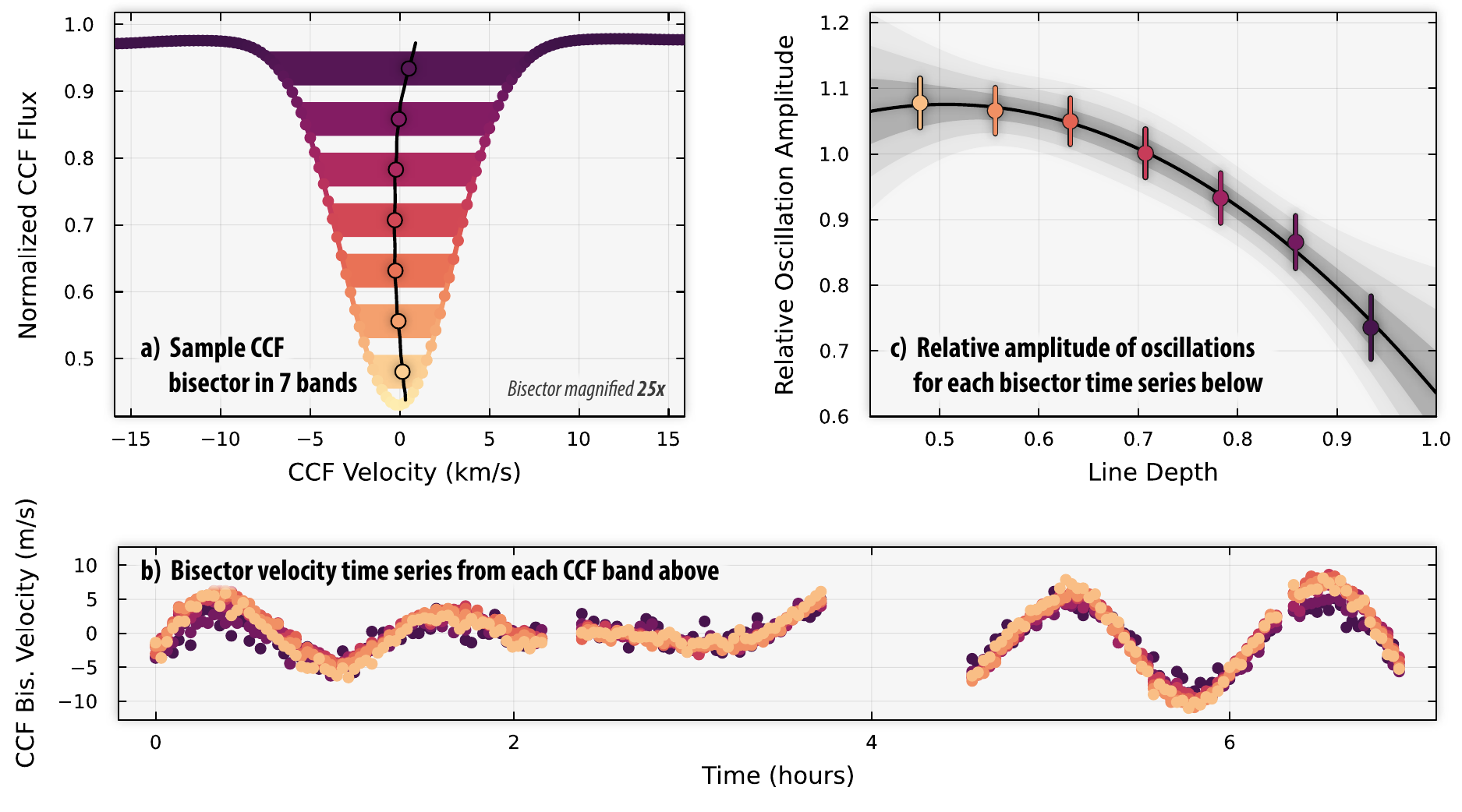}
    \caption{\emph{Top left (a):} Schematic of CCF slices used to compute the bisector velocities in \autoref{sec:bis_velocities}. The CCF bisector is shown, magnified by a factor of 25. \emph{Bottom (b):} Bisector velocity time series for each of the 7 CCF slices shown in panel (a). Points are color coded according to the mean CCF flux level in each slice as in panel (a). The structure is similar across all CCF slices, with a noticeable reduction in amplitude seen for the slices higher in the CCF. \emph{Top right (c)}: Amplitude of oscillations (relative to the oscillation component of the bulk RV time series) for each of the CCF flux bin time series in panel (b). The increase in amplitude for bisector velocities derived closer to the center (core) of the CCF can be interpreted as probing larger velocities due to decreased density higher in the stellar atmosphere. A second-order polynomial is favored over a simple linear fit, shown in black with 1-, 2-, and 3-$\sigma$ contours shown in gray.}
    \label{fig:bisector_velocities}
\end{figure*}
As a final probe of CCF morphology, we compute the CCF bisector velocities within 7 CCF flux bins to diagnose asymmetric shape changes across the CCF.

We break up the CCF into 7 horizontal bands between $0.45 \le F_{CCF} \le 0.98$ and including a gap between bands of 0.025. For each flux band, we calculate the mean velocity of the left (blue) and right (red) wings of the CCF separately, before finally computing the midpoint of the two wings. Velocity uncertainties are estimated from photon-limited velocity uncertainties from CCFs following Equation A.2 of \citet{Boisse2010}, which corrects for oversampling by accounting for the native NEID pixel step size (0.6~\kms\ per pixel) relative to the CCF step size (0.25~\kms). The raw bisector velocities are offset by several hundreds of \ms\ due to the characteristic asymmetry in line profile shapes created by granulation \citep{Dravins1981}; we subtract off the median bisector velocity for each band to show relative bisector velocity variations. The horizontal CCF bands and the resulting bisector velocities of each bands are shown in \autoref{fig:bisector_velocities}.

While the overall structure of the bisector velocity time series looks similar for each CCF band, there is a noticeable reduction in amplitude for the shallowest depth bin. To test the significance of the reduction, we perform the following procedure. We first perform a multi-component  GP fit to each CCF bin's bisector velocity time series as described in \S 4.3.1 of \citet{Gupta2022}. Specifically, we use the GP decomposition methods described in their Equations 19 \& 20 to calculate the predictive means for the individual granulation and oscillation components for each bin's time series\footnote{\S 2.4.5 of \citet{Duvenaud2014} gives a generalized expression for the decomposition of the posteriors of a multi-component GP model. \citet{Aigrain2016} shows an application of this used to separate instrumental systematics from stellar variability in K2 light curves}; we demonstrate the decomposition in \autoref{sec:decomposition}. We then subtract off the granulation component to isolate only the oscillation component and any white noise measurement error.\footnote{In practice this is roughly equivalent to passing the bin's bisector velocity time series through a high-pass filter to remove low-frequency variations.} We fit these residuals to find the amplitude scale factor, $A_i$ that best satisfies
\begin{equation}
 V_{{\rm bis},i}(t) - \mu_{{\rm bis},i}^{\rm{gran}}(t) = A_i  \mu_{RV}^{{\rm osc}}(t),
\end{equation}
where $V_{bis,i} (t)$ is the bisector velocity time series for the $i$\textsuperscript{th} CCF flux bin, $\mu_{bis,i}^{gran}$ is the GP-conditioned mean for the granulation component of $V_{bis,i}$, $A_i$ is the scale factor we fit for, and  $\mu_{RV}^{osc}(t)$ is the GP-conditioned mean for the oscillation component of the pipeline-derived RV time series. In a simpler sense, this can be thought of as fitting each CCF bin's bisector velocity time series as a constant scale factor times the bulk RV time series. However, by removing the granulation component and fitting using the bulk RV oscillation component, we are less susceptible to low frequency variability (i.e., linear and quadratic trends) affecting the final amplitude parameter; our findings are consistent through either method and result in similarly significant linear trends ($>5.5\sigma$). We perform the fit with a single-chain MCMC using the Metropolis Hastings algorithm to sample 100,000 steps; we use a uniform prior on $A_{i}$ between 0 and 2. 

The resulting amplitudes for the 7 CCF flux bins are shown in panel c of \autoref{fig:bisector_velocities}, and show a significant trend\footnote{A quadratic fit is strongly preferred over a flat line (7.2$\sigma$) and over a linear trend (2.6$\sigma$).} of decreasing amplitude for bisector velocities that come from higher in the CCF profile: the oscillation amplitudes from the topmost slice of the CCF (the wings) are 25\% lower than the oscillation amplitudes of the bulk RVs and 31\% lower than the oscillation amplitudes in the bottommost slice of the CCF (the core). We attribute this effect to tracing physical velocities at different heights in the stellar atmosphere, as the CCF is a rough proxy for the average line profile. Because density drops steeply with height in the atmosphere, the velocities higher in the atmosphere---traced by the core of the CCF---are larger than those lower in the atmosphere traced by the wings. This argument has been widely used to explain differences in mode amplitudes (in the Fourier domain) in solar oscillations, first seen with velocities measured simultaneously from Na \textsc{i} and K \textsc{i} absorption lines \citep{Isaak1989}; the difference in species used to measure oscillations is also the reason why solar oscillation velocity amplitudes as measured by GOLF (Na D1/D2) are larger than those measured by BiSON (K) \citep{Baudin2005,Houdek2006,Kjeldsen2008}. 

The limited number of existing solar oscillation instruments, and often shared atomic line species they use, have restricted the detail with which we can probe the dependence of atmospheric height on oscillation amplitude. Our results here suggest a potential avenue for more detailed studies of both solar---through Sun-as-a-Star EPRV efforts, \citep[e.g.,][]{Lin2022,Rubenzahl2023}---and \emph{stellar} p-mode oscillations. Further, we have measured this effect solely in the time domain, thanks to the precision and stability of NEID. A similar analysis in the Fourier domain would be prone to ambiguity due to power leakage from unresolved oscillation modes. It would likely require costly multi-night observing runs to robustly perform this analysis on individually resolved modes. On the other hand, the increased resolution offered by such a study might provide even deeper insights, since different modes probe different regions of the stellar interior.

Of course, our CCF slices are not correlated one-to-one with atmospheric height/density, and so we must treat this interpretation carefully. A more robust analysis would require a detailed spectral synthesis model from which we could derive atomic species (and related line formation physics) and formation heights/atmospheric densities within individual lines as in \citet{AlMoulla2022}. We could then directly compare the measured amplitudes to predictions from theoretical models \citep[e.g.,][]{Zhou2021}. These analyses are beyond the scope of this work, and we leave them to a follow-up study. As an intermediate step, in the next section we perform a line-by-line analysis to provide a more direct view of the effects of p-mode oscillations on spectral line profiles. In particular, we will group lines by their continuum-normalized depth, which avoids the ambiguities imposed by the CCF, which averages over lines of varying depths.

\subsubsection{SCALPELS analysis}
\begin{figure}
    \centering
    \includegraphics[width=\columnwidth]{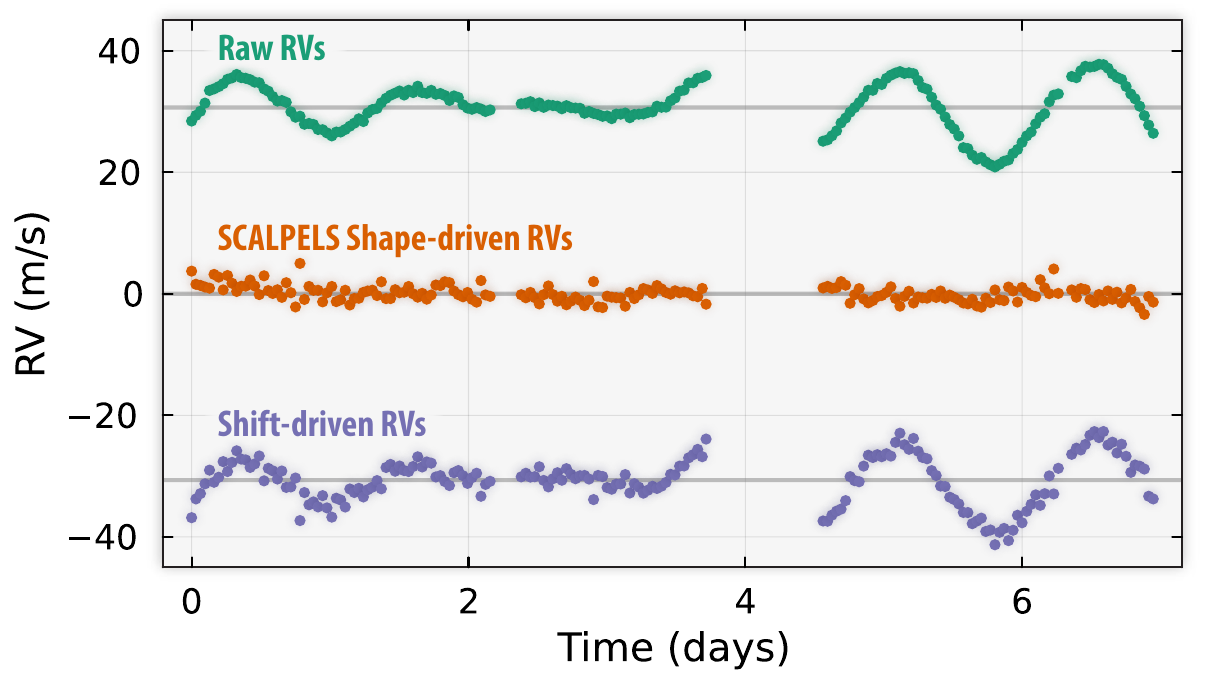}
    \caption{Results of a SCALPELS analysis \citep{CollierCameron2021} to disentangle ``shape-driven" from ``shift-driven" RV variations. The  shape-driven velocities are consistent with a noise-dominated non-detection.}
    \label{fig:scalpels}
\end{figure}

We have shown that the p-mode oscillations in \target\ behave primarily as pure Doppler shifts, where the top of the CCF has a lower amplitude than the bottom. We now wish to test whether these CCF distortions can be recovered with SCALPELS \citep{CollierCameron2021}, an algorithm that leverages the shift-invariant property of the autocorrelation function (ACF) of the CCF to diagnose the presence (and behavior) of potential shape-driven RVs in CCF profiles. We follow the basic SCALPELS procedure as outlined in \citet{CollierCameron2021}. We first compute the 2-dimensional ACF of the CCF profiles (restricted to the portion of the CCF with velocity within $\pm$25~\kms). We compute the singular value decomposition, retaining the first 12 basis vectors (unsorted) after performing the leave-one-out cross validation test in \citet{CollierCameron2021} to determine the number of significant principal components.\footnote{We note that our results are similar if we sort the basis vectors by their response amplitude in the time domain, in which case 4 principal components are deemed significant.} These basis vectors are then multiplied by the original velocities to construct the portion of those velocities due to shape change variations. We then evaluate the ``cleaned", or ``shift-driven" velocities by subtracting out the shape-driven component. The SCALPELS velocities are shown in \autoref{fig:scalpels}. 

We find that the SCALPELS-derived shape-driven component is small compared to the overall amplitude of oscillations in the time series, with an RMS of only 1.29~\ms. Removing this component does not provide a reduction in the overall RMS of the time series, rather it increases slightly from 3.79~\ms\ to 3.83~\ms. The resulting ``cleaned" velocities in \autoref{fig:scalpels} are nearly unaltered and still show clear evidence of p-mode oscillatory behavior. These results indicate that SCALPELS has not detected any shape-driven changes in the CCF profiles.

To test why SCALPELS was unable to detect the CCF morphology changes that we know are present, we perform an injection recovery test with synthesized CCF profiles. We use the following steps to generate pseudo CCFs with ultra-fine resolution:
\begin{enumerate}
    \item Construct a template Gaussian CCF with ultra-fine resolution
    \item Add a common static bisector shape, drawn from the polynomial fit to the average bisector for our observations
    \item Add a time-varying, CCF-flux-dependent bisector: this bisector has velocity equal to the measured RV time series scaled at each CCF-flux level by the amplitude dependence that we measured in the CCF slices (panel c of \autoref{fig:bisector_velocities})
    \item Inject a pure Doppler shift component, synthesized from a sine curve with 50 cm/s amplitude. This ensures that we can recover a pure shift component to check that SCALPELS doesn't attribute all RVs as shape-driven.
\end{enumerate}

These steps are performed on the ultra-fine resolution template CCF to ensure that we don't introduce errors from the interpolation steps when injecting bisectors and pure shifts. We then downsample the CCF to the NEID CCF step size (0.25 km/s) and compare the shape-driven RVs that SCALPELS returns to the known input shape-driven RVs. We repeat this exercise for 30 different CCF white noise amplitudes, drawing 5 white noise realizations for each amplitude level. We define a SCALPELS detection as one where the recovered shape-driven time series is well correlated with the input signal (Pearson correlation coefficient $>0.5$) and the residuals of the shape driven component (recovered shape RVs minus true shape RVs) reduces the RMS by more than 30\% (RMS of the residuals divided by RMS of the shape RVs $< 0.7$). 

We find that the CCF noise level necessary to make consistent detections of shape-driven RVs is 24 ppm. The median RMS of our observed CCFs is 130 ppm, a factor of $\sim$5 larger than this threshold. We conclude that our measured CCFs are too noisy for SCALPELS to discern these shape changes, and thus a non-detection was expected. Despite the large overall amplitude of the oscillations in the RV time series, the shape-change differential is relatively small. In an idealized scenario where the slope is significantly steeper (e.g., the top of the CCF has oscillation amplitude of 0 and the bottom has oscillation amplitude of 1.5), we find that the observed CCF noise level would be sufficient for SCALPELS to detect.

We expect the results here to be applicable to oscillations that are dominated by low-degree modes, given their tendency to behave as pure shifts. It is thus possible that a time series dominated by high-degree non-radial modes would produce asymmetries detectable by SCALPELS, which could be explored with similar time series of \target\ or other stars. However, our results here show that due to the stochastic nature of which modes are excited at any given time, we cannot expect p-modes to \emph{always} induce shape-driven distortions to the CCF.

\subsection{Line-by-line (LBL) analysis}
The findings above suggest that the p-mode oscillations for our observed time series induce primarily symmetric line changes, with higher-order asymmetries due to velocity amplitude variations at different heights in the atmosphere. We now wish to probe these effects at the spectrum level (rather than CCF level) and seek to establish other characteristic diagnostics of p-mode oscillations in RV spectra to inform other approaches to mitigating their effect. Here, we leverage the high per-spectrum SNR of our NEID observations and perform a line-by-line (LBL) analysis. We follow the approach of \citet{Siegel2022}, who compute RVs and line depth for individual lines across the entire spectrum using a template-matching approach; this toolkit is a very close implementation of \citet{Dumusque2018}. The \citet{Siegel2022} algorithm identifies 3442 unique spectral lines; we remove lines with RV uncertainty $> 75$~\ms, which mostly removes the shallowest lines that are likely contaminated by microtellurics. We are left with 2662 lines with median RV uncertainty of 28~\ms\ per line. 

\subsubsection{Line Depth}
\begin{figure*}
    \centering
    \includegraphics[width=\textwidth]{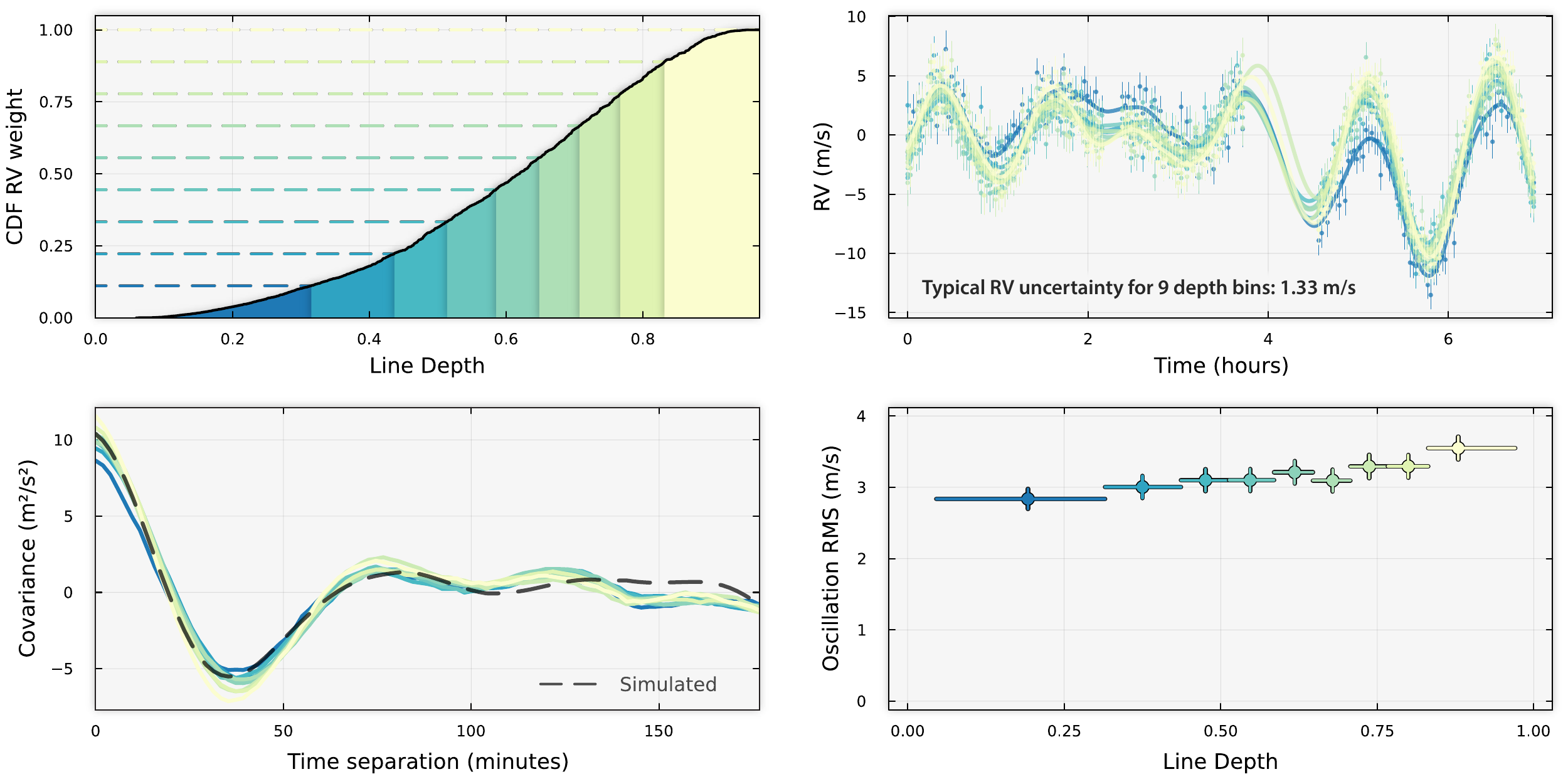}
    \caption{Analysis of line depth dependence of p-mode oscillations for \target. \emph{Top left:} Cumulative distribution function for the relative RV weight of each line as a function of line depth, with depth bins chosen to maintain constant total RV weight for each bin. \emph{Top right:} RV time series for each line bin, with typical RV uncertainty of 1.38~\ms\ per bin; each depth bin's RVs are fit with a two-component GP model. \emph{Bottom left:} Cross-covariance function between the oscillations of each bin and the bulk RVs, where in both cases, we have subtracted out the mean conditioned granulation component from the GP fitting. The dashed line shows a comparison to a simulated time series generated from the same GP model with identical time sampling. There is no measurable phase lag between any depth bins and the bulk RV, nor evidence of a significantly different characteristic timescale. \emph{Bottom right:} Oscillation RMS as a function of line depth. We calculate the RMS of the RV time series shown in the top right panel after subtracting off the GP-conditioned granulation component for each depth bin. The y-errors show the estimated uncertainty on the RMS measurement \citep[see e.g.,][]{Luhn2020a}; the x-errors show the span of each depth bin. There is a noticeable increase in p-mode RMS amplitude for deeper lines; however, the RMS errors are too large for this to be a significant trend. We note that there is no collection of lines that can reduce the oscillation RMS to near 0, indicating that depth-based line selection techniques are not a viable strategy for completely mitigating p-mode oscillations.}
    \label{fig:depth_analysis}
\end{figure*}
First, we investigate whether the p-mode oscillations in our NEID time series of \target\ have a dependence on depth. Our earlier results suggest that such a dependence should exist, with oscillation amplitude increasing with line depth. 

We can now group lines of similar depth and compute the average RV at each average depth. However, shallower lines have lower RV information content (and thus larger RV uncertainties); we therefore construct depth bins that ensure comparable RV uncertainty in the final average RV. To do this, we begin by sorting all 3442 lines by their depth, and calculate the cumulative distribution function of their RV weighting. We then determine the depth spacing that results in even spacing in the RV weight CDF. We choose 9 depth bins\footnote{We note that our results are consistent across choice of binning. Choosing fewer bins naturally results in better RV uncertainty per bin, but at the expense of probing fewer depths; the opposite is true of choosing more bins. We find that 9 strikes a good balance of typical RV uncertainty while probing a wide enough range of line depth.}, and the grouping is shown in the top left panel of \autoref{fig:depth_analysis}. Note that there are many lines contained in the shallowest bin, and fewer lines at the largest depths; the deep-line bins therefore span a wider range of depths.

We then compute the average RV for each depth bin (the typical RV uncertainty per bin is $\sim$1.38~\ms), and perform the same GP modeling to fit each resulting time series, again decomposing the posterior mean into a mean oscillation and mean granulation component. The RVs and total GP mean can be seen in the top right panel of \autoref{fig:depth_analysis}. The RV time series for each depth bin are very similar to both each other and to the bulk RV, though there is a noticeable decrease in amplitude for shallower lines. We note that the shallowest bin has has a slight negative trend over the time series; we observe that this is always true of the shallowest bin no matter how many bins are chosen. We also find that the negative trend is more pronounced when we add back in the lines with RV error $> 75$~\ms, and is in the same direction as the barycentric throw across the observations. This confirms that our line cuts based on RV error per line have successfully removed most microtellurics, though the slight negative trend that remains in the shallow-line bin is evidence that some microtellurics have likely remained in our line list. Since this effectively introduces a linear trend to the RVs, our GP decomposition procedure will attribute such low-frequency variability to granulation. In the calculations below, we focus on only the oscillations by subtracting off the mean granulation component, and thus our results are not affected by  microtelluric contamination.

\begin{figure}
    \centering
    \includegraphics[width=\columnwidth]{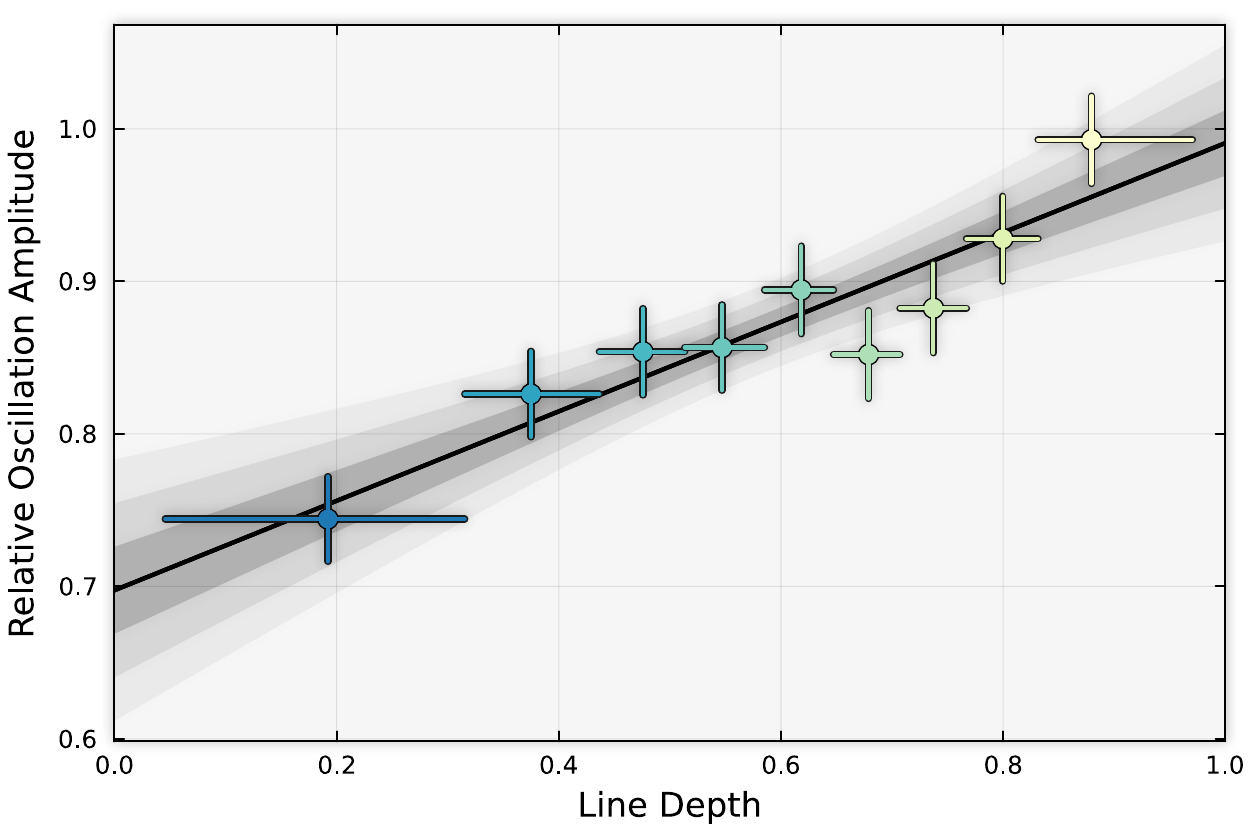}
    \caption{Amplitude of oscillations (relative to the oscillation component of the bulk RV time series) for each depth bin time series in \autoref{fig:depth_analysis}. The increase in amplitude for deeper lines can be interpreted as probing larger velocities due to decreased density higher in the stellar atmosphere.}
    \label{fig:amplitude_depth}
\end{figure}

We first test for any phase lags by calculating the cross-covariance of oscillations in each depth bin against the oscillations of the bulk RV time series, shown in the bottom left panel of \autoref{fig:depth_analysis}; we compute this after subtracting off the mean conditioned granulation components from both the bulk RVs and each depth bin's RVs. We compare these to the covariance structure of a simulated time series generated from the same GP model with identical time sampling. However, since the overall RV time series is relatively short (sampling only $\sim$6 p-mode timescales), any given simulated time series will be prone to stochastic effects, and the observed covariance will differ from the ``true" GP covariance kernels; we therefore generate 100 time series and select the simulated time series that minimizes the residuals to the calculated covariance functions. The cross-covariance structure is largely the same across all depth bins, i.e., there is no measurable phase lag between any depth bins and the bulk RV, nor evidence of a significantly different characteristic timescale. However, we can see evidence of decreased amplitude with shallower depths.   Finally, in the bottom right panel of \autoref{fig:depth_analysis}, we plot the RMS of each depth bin's oscillations, where we see a trend of increasing oscillation RMS for deeper lines, though the RMS errors are too large for it to be a significant trend. Nonetheless, these results echo the earlier findings in \autoref{sec:bis_velocities} that suggest that we are tracing the oscillation velocity profile as a function of atmospheric height, and that oscillations are more vigorous higher in the atmosphere. As a direct comparison, we repeat our amplitude fitting procedure to fit the oscillations of each depth bin as an amplitude scale factor multiplied by the bulk RV oscillations; the amplitudes are shown in \autoref{fig:amplitude_depth}. We see a similar significant trend (6.5$\sigma$), which we again interpret as larger oscillation velocities with increasing height in the stellar atmosphere. However, we repeat the caveat that line depth is not an exact one-to-one substitute for atmospheric height; we encourage a follow-up study that performs the proper spectral synthesis modeling (as described in \autoref{sec:bis_velocities}) to further validate these claims.

\subsubsection{Chromaticity}
We next perform a similar analysis to investigate the potential dependence of p-mode oscillation amplitude on wavelength. For active regions like spots, the contrast between unspotted photosphere and the active region is reduced in the infrared, which leads to a reduced RV amplitude \citep{Reiners2010,Marchwinski2015}. Similarly, p-mode oscillations are detectable in photometry due to their brightness (e.g., temperature) fluctuations, and as such have amplitudes that vary as a function of the bandpass of the observations \citep[e.g.,][]{Lund2019,Sreenivas2025}. RVs directly probe surface velocities, which are not intrinsically chromatic, although we may expect a wavelength dependence arising from spectral line flux in a similar way to photometry. 

\begin{figure*}
    \centering  \includegraphics[width=\textwidth]{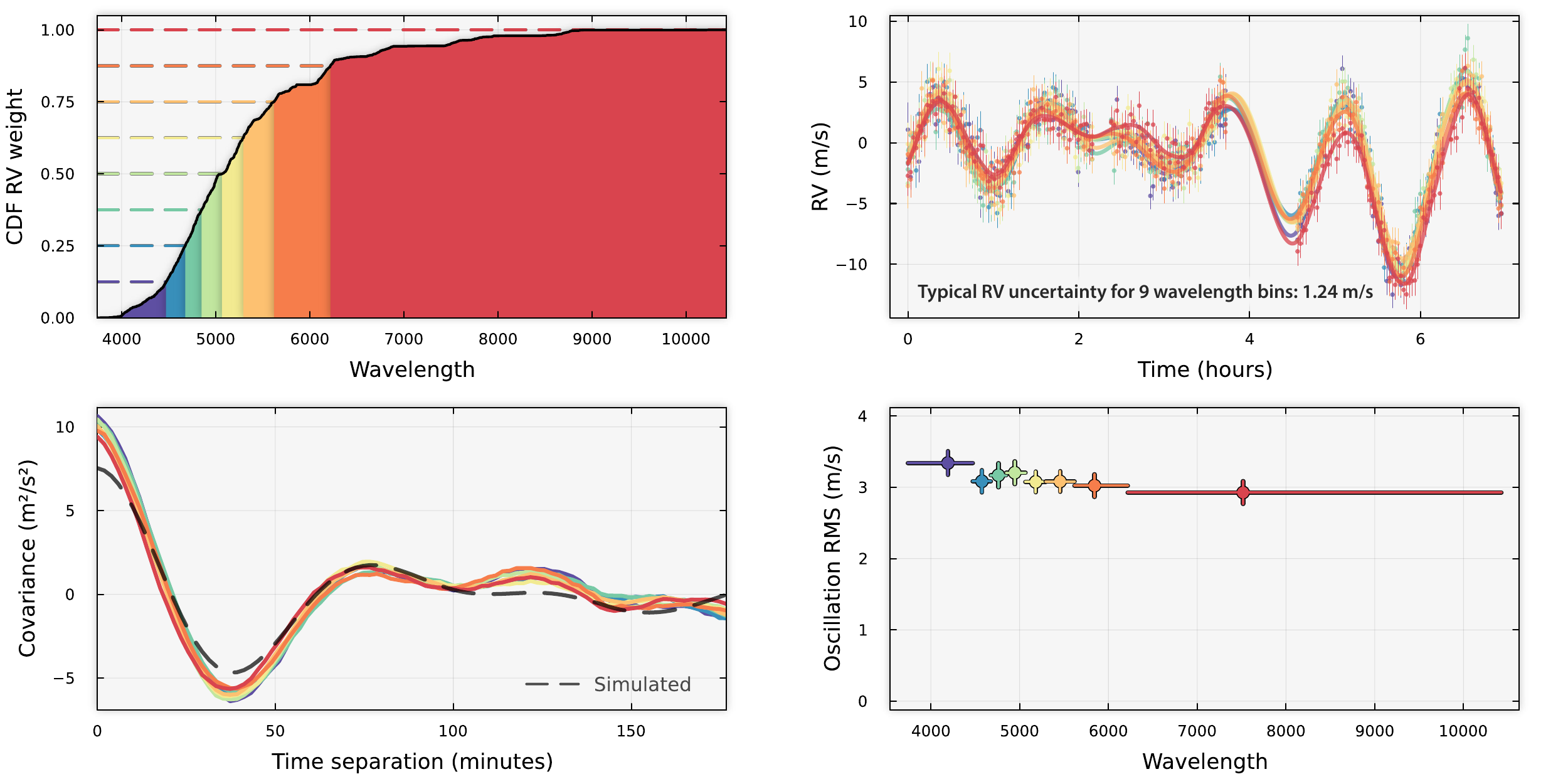}
    \caption{Similar to \autoref{fig:depth_analysis} but for an analysis of chromatic dependence of p-mode oscillations for \target. We again find no evidence of phase lag or a different characteristic timescale for different chromatic bins. There is no significant reduction or amplification in p-mode amplitude at different wavelengths.} \label{fig:chromatic_analysis}
\end{figure*}
We perform the same steps as our depth analysis by: 1) grouping our line list into wavelength bins that preserve equal RV weighting between them, 2) calculating the RV for each wavelength bin and fitting with the two-component GP model, 3) computing the cross-covariance between each wavelength bin's oscillations and the bulk RV oscillations, and 4) plotting the overall oscillation RMS as a function of wavelength for each wavelength bin. These four steps are demonstrated in the four-panel plot of \autoref{fig:chromatic_analysis}.

\begin{figure}
    \centering
    \includegraphics[width=\columnwidth]{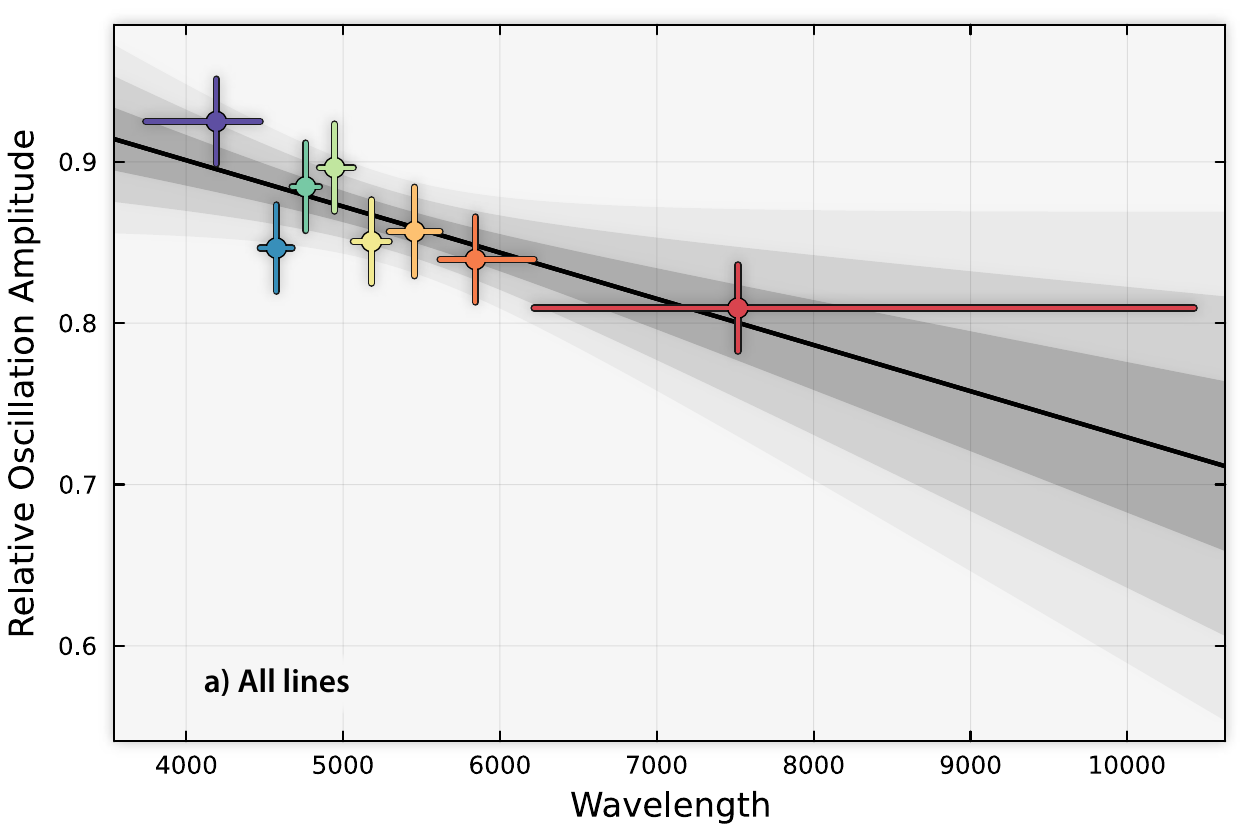}
    \includegraphics[width=\columnwidth]{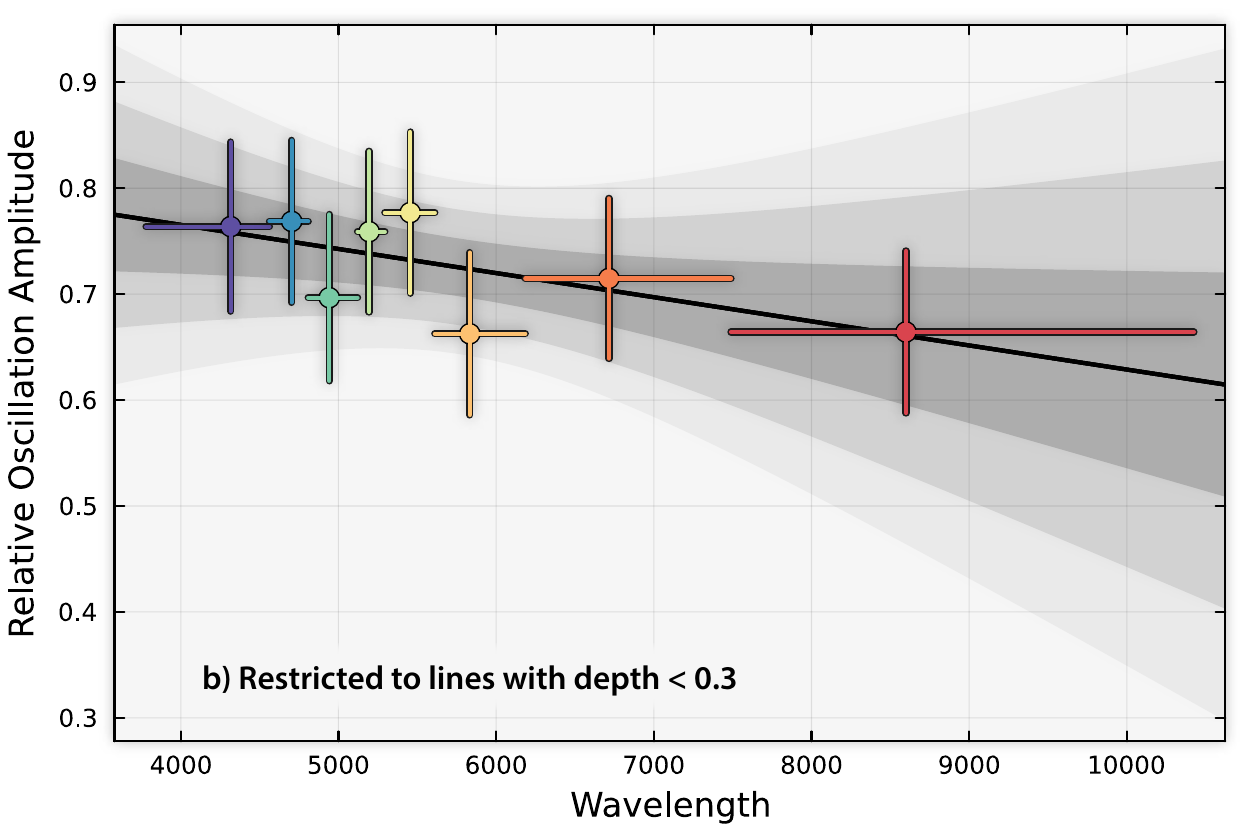}
    \caption{\emph{Top panel:} Amplitude of oscillations (relative to the oscillation component of the bulk RV time series) for each wavelength bin time series in \autoref{fig:chromatic_analysis}. The red-most bin is dominated by shallow lines, and the blue-most bin by deep lines; this change in line depth distribution across the spectrum leads to a marginally significant negative trend (2.9$\sigma$, p-value 0.004). \emph{Bottom panel:} When we restrict the  chromatic analysis to only lines of the same depth ($d <0.3$), the trend is no longer significant (1.1$\sigma$, p-value 0.27).}
    \label{fig:amplitude_wavelength}
\end{figure}

Similar to our depth analysis, we find no evidence of a phase lag or oscillation frequency change as a function of wavelength. While there appears to be a decrease in oscillation RMS with increasing wavelength, this trend is largely driven by the red-most and blue-most bins, and the remaining bins show roughly equivalent oscillation RMS. This is reinforced more clearly in \autoref{fig:amplitude_wavelength}, where we again compute the amplitude scale factors for each wavelength bin relative to the bulk RV oscillations. 
We find a mildly significant wavelength dependence for p-mode oscillation of \target, though the significance of the trend is much lower than the trend seen in the depth analysis ($p$-value 0.004, or 2.9$\sigma$).
The appearance of a trend is likely reflecting the underlying dependence on line depth seen in \autoref{fig:amplitude_depth}, since the line depth distribution changes across the spectrum: the bluest bin contains the deepest lines, and the reddest bin is dominated by shallow lines. Indeed, when we restrict the chromatic analysis to only those lines with line depth $d < 0.3$ so all wavelength bins probe similar line depths (bottom panel of \autoref{fig:amplitude_depth}), the trend is no longer significant (p-value 0.27). 

\section{Discussion}\label{sec:discussion}
\subsection{SCALPELS sensitivity to shape-driven CCF variability}
The measured RV signal due to oscillations was clearly detected in the time series, with RV variations as large as 7 \ms, more than 10 times larger than the estimated single measurement precision. Yet, we were unable to recover this signal as a shape-driven signal in SCALPELS. We caution that these detection limits represent only the detection sensitivity to this specific flavor of variability, where the CCF shifts with an amplitude that depends on the CCF flux level. For variability arising from magnetic activity, e.g., a traveling bump across the CCF profile as a spot rotates into and out of view), the detection sensitivity will be different. Indeed several existing studies have used SCALPELS to identify and remove signals (both instrumental and on rotational timescales) at or below the 1~\ms\ level \citep{CollierCameron2021,John2023,Ford2024}. This highlights the need for more diverse and detailed tests of how various mitigation strategies respond to the presence of different types of stellar variability across a range of timescales.

\subsection{Implications for RV Surveys of Main Sequence Stars}
The observed CCF profile changes induced by p-mode oscillations are a relatively small effect---despite the large amplitude signal in the RV time series, we could not recover it with our SCALPELS analysis. This suggests that detecting such changes for Sun-like stars, where the RV amplitudes are several times smaller, would be even more challenging. While it is possible that the slope of the amplitude change in \autoref{fig:bisector_velocities} could be steeper for main sequence dwarfs with thinner envelopes (and thus the CCF profile may trace a sharper density gradient), we expect that these amplitude differences would still prove to be below the noise floor for a SCALPELS-like detection. This is made worse by the fact that the typical exposure times necessary for high SNR spectra for medium-sized telescopes like NEID are similar to the p-mode timescales for Sun-like stars, thus acting to blur out the signal. In fact, the current typical approach to mitigating the effect of p-mode oscillations in RV planet-search surveys of Sun-like stars is to expose over an integer number of p-mode cycles \citep{Chaplin2019}, and we can expect this method to sufficiently blur the amplitude differences across the CCF bisector to well below the detection threshold.

For all the reasons above, if one is following a \citet{Chaplin2019} approach to mitigating p-mode oscillations (typically down to the 10~\cms\ level), methods like SCALPELS should not be expected to remove any additional residual effects from p-mode oscillations that may still be present in the time series.

\section{Summary and Conclusions}\label{sec:conclusions}
Here we have investigated the impact of p-mode oscillations on RV spectra and related data products. We obtained a 7-hour time series of resolved p-mode oscillations on subgiant \target\ with the NEID spectrograph, with residual RMS of 25~\cms\ after fitting with a oscillation and granulation GP model. Our choice of subgiant target ensured that p-mode oscillations would be easily resolvable given the large (3--5~\ms) amplitudes and long (1.5~hour) periods. We performed several analyses to investigate how p-mode oscillations manifest in RVs.

First, we investigated the nature of CCF morphology changes that p-mode oscillations induce. By examining the residuals to a mean ``template" CCF, we found that p-mode oscillations can be modeled primarily by pure translational Doppler shifts to the CCF profile, indicating a time series dominated by low-degree oscillation modes. Importantly, our results show that we cannot expect p-modes to always induce shape-driven distortions to the CCF.
We further investigated CCF morphology by computing CCF bisector velocity time series for 7 horizontal flux ``slices" from the CCF. By modeling the resulting CCF bisector velocity oscillation time series as an amplitude scale factor multiplied by the bulk RV oscillation time series, we find that the oscillation amplitude varies across the CCF bisector. Specifically, we measure 10\% larger oscillation amplitudes for CCF bisector velocities derived closer to the core of the CCF, and 25\% smaller oscillation amplitudes for bisector velocities derived from the wings of the CCF. We interpret this result as tracing the physical oscillation velocities as a function of height in the stellar atmosphere.

We then performed a SCALPELS analysis \citep{CollierCameron2021}, decomposing the RV time series into ``shift-driven" and ``shape-driven" components and found that SCALPELS was unable to attribute any significant variability due to CCF shape changes. Injection tests with pseudo CCFs revealed that a SCALPELS non-detection was expected given the typical CCF noise level and the relatively small amplitude differences between the top and bottom of the CCF. These results support the conclusion that p-mode oscillations in \target\ manifest primarily as pure Doppler shifts, providing context for the applicability of mitigation techniques like SCALPELS for p-mode oscillations. Extending to Sun-like targets of EPRV-era surveys with weaker oscillations, we should not expect methods like SCALPELS to provide significant oscillation mitigation.

Next, we performed a line-by-line analysis to probe other diagnostics and possible mitigation strategies for p-mode oscillations. We looked first at the dependence of p-mode oscillation amplitude as a function of line depth and found a significant trend of increasing oscillation amplitude for deeper lines, further supporting our previous conclusion that we are probing the oscillation velocity field as a function of stellar atmospheric height. We did not find any evidence of time lags or frequency dependence on line depth. Finally, similar analyses of wavelength dependence demonstrated that there is no significant reduction in p-mode amplitude, nor evidence of time lags or oscillation frequency changes as a function of wavelength. 

We recommend more detailed spectral synthesis analyses on these data as follow-up studies to properly compute formation height/stellar densities as a validation of the physical interpretation of the oscillation amplitude dependence we see. We further encourage similar studies on other targets to investigate how these trends vary as a function of spectral type or evolutionary stage.

\begin{acknowledgements}

J.K.L. is supported by an appointment to the NASA Postdoctoral Program at the NASA Jet Propulsion Laboratory, administered by Oak Ridge Associated Universities under contract with NASA.

This work is based on observations at Kitt Peak National Observatory, NSF’s NOIRLab (Prop. ID 2022A-407334; PI: J. Luhn), managed by the Association of Universities for Research in Astronomy (AURA) under a cooperative agreement with the National Science Foundation. The authors are honored to be permitted to conduct astronomical research on Iolkam Du’ag (Kitt Peak), a mountain with particular significance to the Tohono O’odham.
 
 Data presented herein were obtained at the WIYN Observatory from telescope time allocated to NN-EXPLORE through the scientific partnership of the National Aeronautics and Space Administration, the National Science Foundation, and the National Optical Astronomy Observatory.

We thank the NEID Queue Observers and WIYN Observing Associates for their skillful execution of our NEID observations.

S.M.\ is the NEID Principal Investigator.
P.R.\ serves as NEID Instrument Team Project Scientists.  

C.F.B., A.F.G., S.H., S.M., J.P.N., J.R., P.R., A.R., C.S., G.S., R.C.T, and J.T.W are project architects that made significant and foundational contributions to NEID.

The research was carried out, in part, at the Jet Propulsion Laboratory, California Institute of Technology, under a contract with the National Aeronautics and Space Administration (80NM0018D0004) and funded through the President’s and Director’s  Research \& Development Fund Program.

The Center for Exoplanets and Habitable Worlds 
is supported by Penn State and the Eberly College of Science.

T.R.B.\ acknowledges support from the Australian Research Council through Laureate Fellowship FL220100117.

This research has made use of the SIMBAD database, operated at CDS, Strasbourg, France, and 
NASA's Astrophysics Data System Bibliographic Services.

This research has made use of the NASA Exoplanet Archive, which is operated by the California Institute of Technology, under contract with the National Aeronautics and Space Administration under the Exoplanet Exploration Program.

\end{acknowledgements}

\newpage

\facilities{WIYN (NEID), Exoplanet Archive}

\bibliography{library}

\appendix

\section{GP Decomposition}\label{sec:decomposition}
\begin{figure}
    \includegraphics[width=\columnwidth]{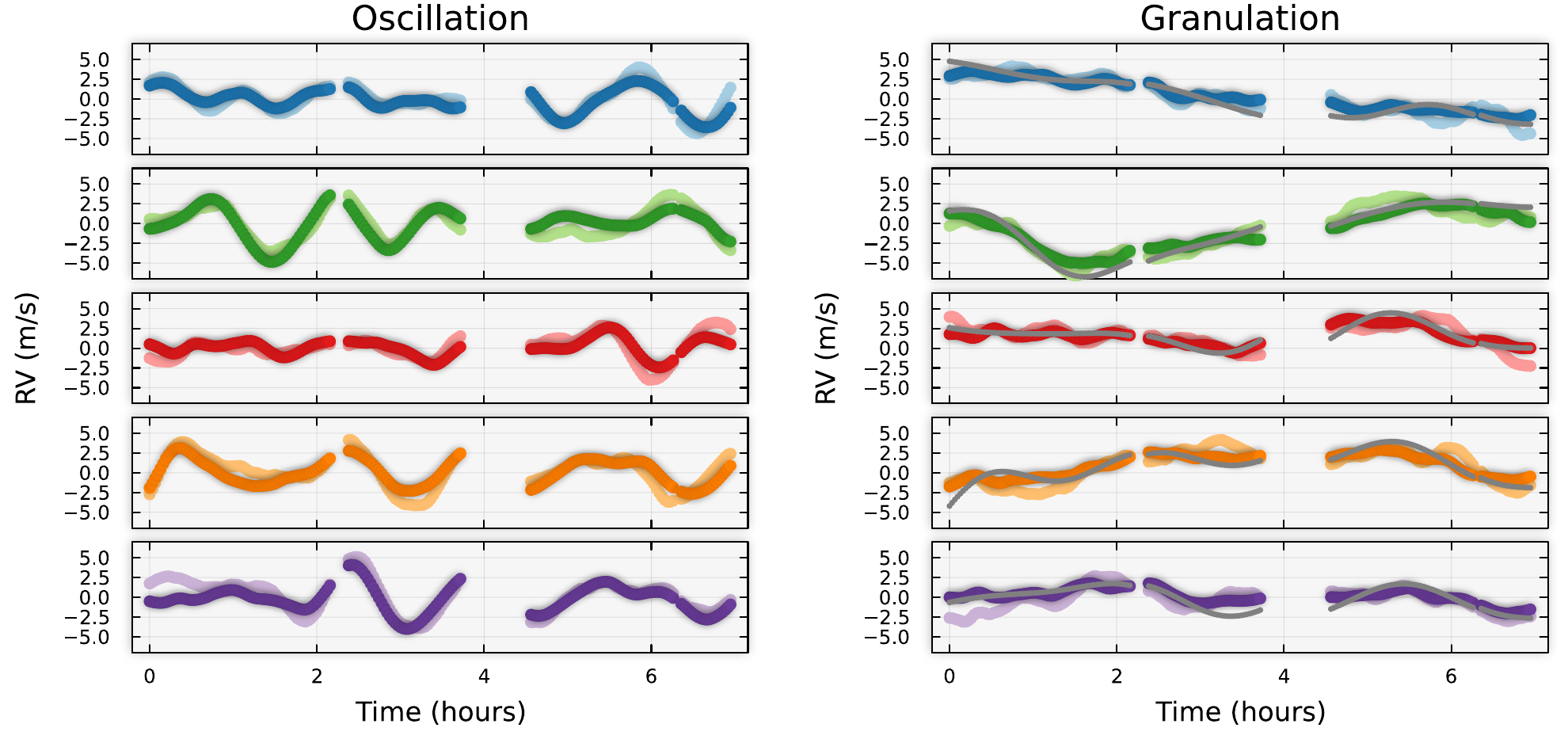}
    \caption{Five randomly selected representative samples demonstrating the GP decomposition procedure. Left columns show the recovered oscillation signal (dark bold colors) compared to the true synthesized signal (light faded colors) for each sample. The panels in the right column show the same for the granulation signal and also show in gray the result of passing the full synthetic time series through a low-pass filter.}
    \label{fig:decomposition_representative_sample}
\end{figure}

\begin{figure}
    \includegraphics[width=\columnwidth]{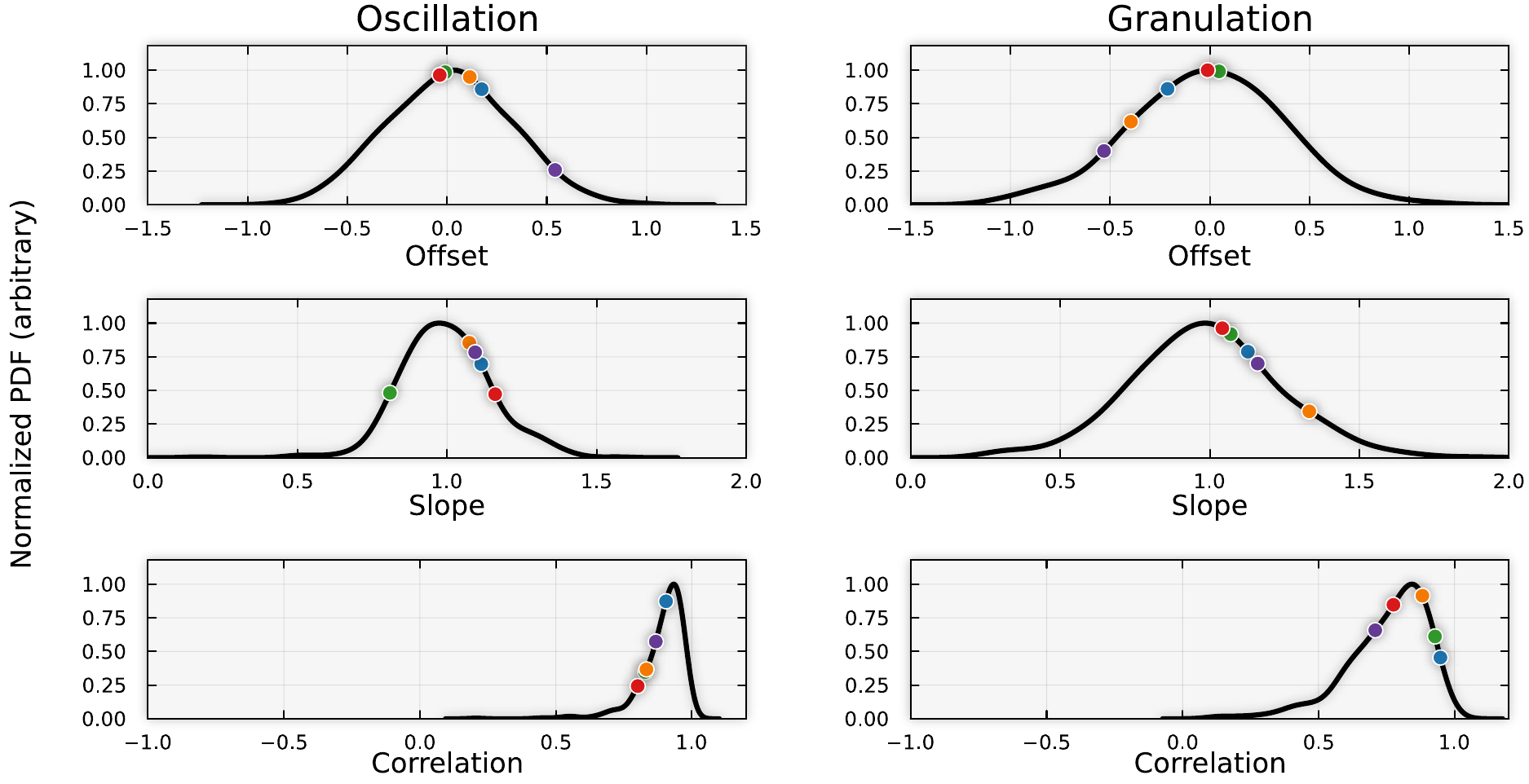}
    \caption{Distribution of summary statistics for the GP decomposition analysis. For the decomposed oscillation (left) and granulation (right) signals, the top two rows show the distributions of the offset (top row) and slope (middle row) of a linear fit to the true synthesized signal, and the bottom row shows the Pearson correlation coefficient between the true and recovered signals. The representative samples from \autoref{fig:decomposition_representative_sample} are plotted at their respective values along the distribution.}
    \label{fig:decomposition_distributions}
\end{figure}

We include here a visual aid for validating the GP decomposition approach used in this work. We simulate 500 ``true" time series for both oscillation and granulation. We add these together with 500 realizations of white noise drawn from the reported photon noise uncertainties to obtain 500 total test time series. For each time series, we compute the decomposition, and extract the predictive means for the oscillation and granulation components individually. A representative sample is shown in \autoref{fig:decomposition_representative_sample}. We then compute the correlation between the predicted and ``true'' synthetic signals, and perform a linear fit between them to obtain the slope and offset. The resulting distributions indicate successful prediction of the individual components (\autoref{fig:decomposition_distributions}). On average, the predicted time series has a slope of 1, with 0 offset, and is strongly correlated with the input ``true" time series (correlation coefficient near 1). In particular, the predicted oscillation signal has stronger correlation with the ``true" signal than granulation, since the kernel amplitude is larger.

As a general rule, this GP decomposition technique will be most successful when predicting constituent components that operate at distinct frequencies with little overlap. In our case, the granulation kernel we use does have appreciable power at the frequencies in the oscillation Gaussian power envelope, though they are significantly lower power than the oscillation kernel at these frequencies. As a result, granulation variability at these frequencies will tend to be attributed to the oscillation component. We indeed see that the predicted oscillation signal tends to differ from the ``true" signal when the ``true" granulation signal shows variability over an oscillation timescale.  That said, the majority of the power in granulation occurs at the lowest frequencies, which means the granulation component is very similar to passing the time series through a low-pass filter. \autoref{fig:decomposition_representative_sample} includes a comparison between the granulation component and a low-pass-filtered version of the input time series, using a cutoff frequency 0.75 times the $\nu_{max}$ frequency. If we use a larger cutoff frequency, the low-pass filter begins to absorb oscillations, highlighting that the GP decomposition technique can better handle components with overlapping power. The GP decomposition is also superior for its ability to naturally handle time series with gaps or unevenly spaced observations.

\end{document}